\documentstyle[12pt, epsfig]{article}
\newcommand{\bea}   {\begin{eqnarray}}
\newcommand{\eea}   {\end{eqnarray}}
\begin{document}
\renewcommand{\thefootnote}{\fnsymbol{footnote}}

\thispagestyle{empty}
\title{One-dimensional $\sigma$-models with \\$N=5,6,7,8$ off-shell supersymmetries}
\author{M. Gonzales\thanks{{\em e-mail: marcbino@cbpf.br}},
M. Rojas\thanks{{\em e-mail: mrojas@cbpf.br}}
~and F. Toppan\thanks{{\em e-mail: toppan@cbpf.br}}
~\\~ \\
{\it $~^{\ast\ddagger}$CBPF, Rua Dr.}
{\it Xavier Sigaud 150,}
 \\ {\it cep 22290-180, Rio de Janeiro (RJ), Brazil.}\\
 {\it $~^\dagger$DEX, Universidade Federal de Lavras, CP 3037,}\\{\it 32700-000, 
 Lavras (MG), Brazil.}}
\maketitle
\begin{abstract}
We computed the actions for the $1D$ $N=5$ $\sigma$-models with respect to the two inequivalent 
$(2,8,6)$ multiplets. $4$ supersymmetry generators are manifest, while the constraint 
originated by imposing the 
$5$-th supersymmetry automatically induces a full $N=8$ off-shell invariance. The resulting 
action 
coincides in the two cases and corresponds to a conformally flat $2D$ target satisfying a 
special 
geometry of rigid type. \par
To obtain these results we developed a computational method (for {\em Maple 11}) which does not 
require the notion of superfields and is instead  based on the nowadays available
list of the inequivalent representations of the $1D$ $N$-extended supersymmetry. Its application 
to systematically analyze the $\sigma$-models off-shell invariant actions for the remaining 
$N=5,6,7,8$ $(k,8,8-k)$ multiplets, as well as for the $N>8$ representations,
only requires more cumbersome computations.
\end{abstract}
\vfill
\rightline{CBPF-NF-026/08}

\newpage
\section{Introduction}
In recent years the representation theory of the $1D$ $N$-extended supersymmetry algebra has 
been substantially clarified, see \cite{pt,krt,dfghil,kt,kt2}. Its knowledge has been used to
investigate $N$-extended $1D$ $\sigma$-models, both for linear \cite{bikl}, as well as for 
non-linear realizations (see e.g. \cite{nlin}) of the extended supersymmetry. The main, but 
not exhaustive reference for the linear $N=8$ $1D$ $\sigma$-models is \cite{bikl}
(further developments have also been investigated in \cite{ils}). An $N=8$-superfield 
description was given for the whole list 
of $(k,8,8-k)$ multiplets
(where $k=0,1,2,\ldots, 8$ is the number of coordinates of the target manifold, $8$ 
is the number of fermionic fields, while $8-k$ is the number of auxiliary fields). 
A corresponding list of $N=8$ off-shell invariant actions was also produced. 
The \cite{bikl} results, however, are not complete for two independent reasons. 
The $N=8$ off-shell actions there listed are not the most general ones 
(at least for some values of $k$). Furthermore, no discussion was made
concerning a non-maximal number of extended supersymmetries, namely the most general 
off-shell invariant actions associated to $N=5,6,7$ supersymmetry generators 
(they are also irreducibly represented in the $(k,8,8-k)$ multiplets; on the other hand 
$N=8$ is the maximal number of supersymmetry generators that such multiplets 
carry). For what concerns the first point, in \cite{krt} it was produced the 
most general $N=8$ off-shell action for $k=1$ (the $(1,8,7)$ multiplet), later 
studied also in \cite{di}. The second point is very specific 
of the limitations of the superfield formalism, not very suitable to deal with such 
issues (no $N=5,6,7$ superfield formalism is available). The situation is further 
complicated by the \cite{dfghil} observation that inequivalent supersymmetry transformations 
acting on the same field-content (the $(k,8,8-k)$ multiplets for a fixed $k$) can be encountered. In \cite{kt} the classification of such 
cases (for $N\leq 8$ they only exist for $N=5,6$) was given in terms of the inequivalent 
connectivities of the graphs associated to the supersymmetry transformations.
In \cite{kt2} it was further proven that the $N=5$ inequivalent connectivities are 
in consequence of the different decompositions in terms of the $N=4$ subalgebras.
In this work we investigate for the first time the question whether the inequivalent $N=5$ supersymmetry transformations induce inequivalent $N=5$ off-shell actions. \par
We developed a general method (concretely implemented in a computational package for Maple $11$, a version which allows performing algebraic computations with grassmann fields). Our framework allows to impose constraints, in sequence, arising from the $5$-th, $6$-th, $7$-th and $8$-th supersymmetry,
studying whether an $N=5$ off-shell invariance automatically induces an $N=8$ invariance. This issue
is related with deep properties of supersymmetry. In \cite{bbm} it was shown that, under a twist,
the action of a $D=4$ $N=2$ Super-Yang-Mills is determined by $5$ of its $8$ supersymmetry generators. A $D=4\rightarrow D=1$ dimensional reduction of these systems produces an $N=8$ Supersymmetric Quantum Mechanics (such that a vector multiplet $(3,4,1)$ and a matter multiplet $(2,4,2)$ are combined in a single $(5,8,3)$ $N=8$ multiplet, see \cite{top}). The class of $N=5$ supersymmetric theories is more general than the ones obtained from $D=4$ dimensional reduction of SYM theories. It is quite important to investigate whether the \cite{bbm} result is in consequence of some non-trivial deep property of the extended supersymmetry.\par
The computational method discussed in the next Section is a generalization of the one used in \cite{krt} to derive the $(1,8,7)$ $N=8$ off-shell action. The problem in \cite{krt} was considerably simplified by the symmetry properties of the $(1,8,7)$ multiplet, which can be covariantly expressed in terms of the octonionic structure constants. For more general multiplets, not exhibiting special symmetry properties, a brute-force technique is more suitable. We concretely applied it to the $k=2$ multiplets ($(2,8,6)_A$ and $(2,8,6)_B$, see Appendix). More cumbersome computations can be carried on for $k>2$ and $N=5,6,7$ length-$4$ multiplets (see \cite{kt}). In the next Section we describe the
mathematical framework. The concrete application to the $N=5,6,7,8$ one-dimensional $\sigma$-model
with $(2,8,6)$ field content will be discussed in Section {\bf 3}. The resulting action is given in Section {\bf 4}. In the Conclusions we discuss our results and point out possible applications
and further lines of development. To save space we invite the reader to consult the references \cite{kt} and \cite{kt2} 
for the information concerning the graphical interpretation of the supersymmetry transformations, the notion of multiplets of different length and/or inequivalent connectivity. The needed supersymmetry transformations are explicitly given in the Appendix.

\section{$N$-Extended off-shell actions.}

We discuss a method to explicitly construct one-dimensional actions which are $N=5,6,7,8$ off-shell invariant under the supersymmetry transformations closing the
\bea
\{Q_i,Q_j\}&=& H,\nonumber\\
\relax [H, Q_i]&=& 0
\eea
$1D$ $N$-extended supersymmetry algebra (where $i,j=1,2,\ldots, N$, while $H$ coincides with a time-derivative).
\par
As discussed in the Introduction we are dealing with multiplets with $(k,8,8-k)$ field-content, given by $k$ fields of mass-dimension $0$ (interpreted as coordinates of a $k$-dimensional target manifold),
$8$ fermions of mass-dimension $\frac{1}{2}$ and $8-k$ auxiliary fields of mass-dimension $1$.\\
A lagrangian ${\cal L}$ generating no higher-derivatives equations of motion has mass-dimension $2$ (the action is ${\cal S}=\frac{1}{m}\int dt {\cal L}$).\par
For any $k=1,2,\ldots, 8$, the most general homogeneous term ${\cal T}_d$ of mass-dimension $d$, constructed with the fields entering the $(k,8,8-k)$ multiplet and their time-derivatives (a time-derivative counts as $1$ in mass-dimension), involves the following number of independent functions of the $k$ target coordinates:  
\bea
{\cal T}_0&:& 1 \quad function,\nonumber\\
{\cal T}_\frac{1}{2}&:& 8 \quad functions,\nonumber\\
{\cal T}_1&:& 36 \quad function,\nonumber\\
{\cal T}_\frac{3}{2}&:& 128 \quad functions,\nonumber\\
{\cal T}_2&:& 402 \quad functions,\nonumber\\
{\cal T}_\frac{5}{2}&:& 1152 \quad functions.
\eea
Collectively denoting with ${\vec x}$ the $k$ target coordinates and with $\psi_j$ the $8$ fermionic fields we have, e.g., that ${\cal T}_0 \equiv F({\vec x})$, ${\cal T}_\frac{1}{2}\equiv \sum_jF_j({\vec x})\psi_j$ and so on.\par
A manifestly  ${\overline N}$-extended supersymmetric lagrangian ${\cal L}_{\overline N}$ is produced through the position 
\bea
{\cal L}_{\overline N} &=& Q_1\cdots Q_{\overline N} F_{\overline N},
\eea
with, in mass-dimension, $\relax [ {\cal L}_{\overline N}] = 2$, $\relax [ F_{\overline N}]=  2-\frac{\overline N}{2}$,
provided that ${\cal L}_{\overline N}$ is not expressed as a time-derivative of some $d=1$ function. \par
The supersymmetry operators $Q_i$ have mass-dimension $[Q_i]=\frac{1}{2}$, making clear that we can have at most ${\overline N}=4$ manifest supersymmetries, with a lagrangian expressed in terms of unconstrained functions entering $F_{\overline N}$ (their total numbers are $1,8,36,128,402$ for ${\overline N}=0,1,2,3,4$, respectively).\par
In order to have an $N$-extended supersymmetric action, with $N>{\overline N}$, we have to impose
$N-{\overline N}$ constraints, for $j={\overline N}+1, \cdots, N$, expressed through
\bea\label{constraints}
Q_j{\cal L}_{\overline N} &=& \partial_t R_{j,{\overline N}},
\eea
where, in mass-dimension, we have
\bea
\relax [R_{j,{\overline N}}]&=&\frac{3}{2}.
\eea
Each (\ref{constraints}) constraint generates a system
of $1152$ constraining equations to be solved in terms of $128$ functions (the coefficients entering $R_{j,{\overline N}}$). Needless to say, the great majority of the $1152$ equations are trivially satisfied, while most of the remaining ones are redundant, generating some constraint which is repeated over and over.\par
This is a general scheme that, with straightforward modifications, can be applied to $N>8$ extended supersymmetries and multiplets of length greater than $3$. In the next Section  we concretely apply it to analyze the off-shell invariant actions of the $k=2$ $(2,8,6)$ multiplets.

\section{The $N=5,6,7,8$ $\sigma$-model with $(2,8,6)$ field-content.} 

There are two inequivalent $N=5$ multiplets with $(2,8,6)$ field-content \cite{dfghil,kt,kt2}. 
The first one, $(2,8,6)_A$, is given \cite{kt2} by the $N=4$ (the supersymmetric operators ${\widehat Q}_1,{\widehat Q}_2,{\widehat Q}_3,{\widehat Q}_4$ in the Appendix)
irreducible multiplets $(2,4,2)$ and $(0,4,4)$ linked together by a $5$-th supersymmetry (the operator ${\widehat Q}_5$). The second one, $(2,8,6)_B$, is given \cite{kt2} by two $N=4$ (the supersymmetric operators
${\overline Q}_1,{\overline Q}_2,{\overline Q}_3,{\overline Q}_4$ in the Appendix) irreducible multiplets 
$(1,4,3)$ linked together by a $5$-th supersymmetry (the operator ${\overline Q}_5$).
\par
In the latter case we can write a manifestly $N=4$ lagrangian ${\cal L}$, written as
\bea\label{N4}
{\cal L} &=& {\overline Q}_1{\overline Q}_2{\overline Q}_3{\overline Q}_4 F(x,y),
\eea
with $F$ an unconstrained function of the two target coordinates. Imposing, according to (\ref{constraints}),
the invariance of the action under the $5$-th supersymmetry ${\overline Q}_5$ has the effect of constraining $F(x,y)$, which must satisfy 
\bea\label{constr}
\Box F &=& \partial_x^2F +\partial_y^2 F =0.
\eea
No further constraint arises by imposing the $6$-th, $7$-th or $8$-th supersymmetry (the operators
${\overline Q}_6,{\overline Q}_7,{\overline Q}_8$, respectively). Therefore, the invariance under $N=5B$ automatically induces the invariance under the full $N=8$ extended supersymmetry.\par
For what concerns the first case, it is a priori unclear whether a manifestly $N=4$ invariant lagrangian can even be produced. The $\left(
\begin{array}{c}
8\\
4
\end{array}\right)=70
$ inequivalent choices of $4$ supersymmetry operators ${\widehat Q}_i$ out of the total set of $8$ operators
produce the following results: in $16$ cases a manifest $N=4$ lagrangian involving all fields entering the $(2,8,6)_A$ multiplet can be encountered; in $6$ cases the $N=4$ lagrangian involves only the fields entering a $(2,4,2)$ submultiplet (due to the fact that there are $6$ inequivalent ways of selecting $N=4$-closed
$(2,4,2)$ multiplets inside $(2,8,6)_A$); the remaining $48$ cases give a total time-derivative.  
\par
We should not accept for granted that, starting from a manifestly $N=4$ lagrangian, we can get the most general off-shell invariant action for $N>4$ with the procedure discussed in the previous section. We explicitly verified the construction starting from a manifestly $N=3$ lagrangian which, in principle, can
provide a more general result. After setting
\bea
{\cal L} &=& {\overline Q}_1{\overline Q}_2{\overline Q}_3(\sum_j F_j(x,y)\psi_j)
\eea
we imposed the set of constraints generated by $Q_4$ and, independently, the set of constraints generated by $Q_5$.\par
The $Q_4$ constraints give the equations
\bea\label{q4}
\partial_x F_7&=&-\partial_y F_8,\nonumber\\
\partial_x F_8 &=& \partial_y F_7,\nonumber\\
\partial_x F_6&=&\partial_y F_5,\nonumber\\
\partial_x F_3 &=& \partial_y F_4.
\eea
The
$Q_5$ constraints give the equations
\bea\label{q5}
\partial_x F_7&=&-\partial_y F_8,\nonumber\\
\partial_x F_8 &=& \partial_y F_7,\nonumber\\
\partial_x F_6&=&\partial_y F_5,\nonumber\\
\partial_x F_1 &=& -\partial_y F_2
\eea
(the only new equation is the last one).\par
The second, third and fourth equations in (\ref{q4}) and (\ref{q5}) can be solved in terms of the unconstrained 
fields $H_1, H_2, H_3,H_4$ s.t. 
\bea
F_7=\partial_x H_1, && F_8=\partial_y H_1, \nonumber\\
F_5=\partial_x H_2, && F_6=\partial_y H_2, \nonumber \\
F_4=\partial_x H_3, && F_3=\partial_y H_3, \nonumber\\
F_2=\partial_x H_4, && F_1=-\partial_y H_4. 
\eea
The first equation in (\ref{q4},\ref{q5}) produces the constraining equation  $\Box H_1=0$. \par
No further constraint arises by imposing invariance of the action under ${\widehat Q}_6$, ${\widehat Q}_7$
or ${\widehat Q}_8$.
\par 
The terms in the lagrangian depending on $H_2,H_3,H_4$ correspond to a total derivative and can be 
eliminated. For instance, setting for simplicity $K\equiv H_3$, its corresponding term reads as
$\partial_t( K_xg_4-K_yg_3-K_{xx}\psi_5\psi_2-K_{xy}\psi_1\psi_5+K_{xy}\psi_2\psi_6-K_{yy}\psi_1\psi_6)$.
The action only depends on the constrained field $H_1$. It coincides with the action produced from a manifestly $N=4$ construction and is automatically $N=8$ invariant. Since, contrary to the $N=5$ transformations, the $N=8$ transformations are uniquely defined \cite{kt}, the action coincides with the one obtained by imposing the $N=5B$ invariance.\par
The overall result can be summarized in the following statements:\\
- the $N=5A$ invariance automatically induces an $N=8$ off-shell invariant action;\\
- the $N=5B$ invariance automatically induces an $N=8$ off-shell invariant action;\\
- the $N=8$ invariant action, uniquely determined in terms of a single, constrained function satisfying (\ref{constr}), is also invariant under the two inequivalent $N=5$ and the two inequivalent $N=6$ supersymmetry transformations.

\section{The action} 

The most general off-shell invariant action is expressed in terms of a single prepotential
$\Phi(x,y)$ satisfying the constraint
\bea\label{constr2}
\Box \Phi &=& \partial_x^2\Phi +\partial_y^2\Phi =0
\eea
($\Phi$ is explicitly recovered in terms of $F$ entering (\ref{N4}) as $\Phi=\partial_x^2F$).
\par
The lagrangian of the system is explicitly given by (the summation over repeated indexes is understood)
\begin{eqnarray}\label{lagr}
\relax {\cal L}&=& \Phi(\dot{x}^{2}+\dot{y}^2 - \psi_{0}\dot{\psi}_{0}
-\psi_{i}\dot{\psi}_{i}-
\lambda_{0}\dot{\lambda}_{0}
-\lambda_{i}\dot{\lambda}_{i}+ g_{i}g_{i}+f_{i}f_{i})+\nonumber\\
&& +\Phi_x[\dot{y}(\psi_0\lambda_0-\psi_i\lambda_i)-g_i(\psi_i\psi_0+
\lambda_i\lambda_0)+
f_i(\psi_i\lambda_0-\lambda_i\psi_0)+\nonumber\\
&&\epsilon_{ijk}(f_i\lambda_j\psi_k+\frac{1}{2}g_i(\lambda_j\lambda_k-\psi_j\psi_k))]+\nonumber\\
&&
-\Phi_y[\dot{x}(\psi_0\lambda_0-\psi_i\lambda_i)+f_i(\psi_i\psi_0+
\lambda_i\lambda_0)+
g_i(\psi_i\lambda_0-\lambda_i\psi_0)+\nonumber\\
&&-\epsilon_{ijk}(g_i\psi_j\lambda_k-\frac{1}{2}f_i(\lambda_j\lambda_k-\psi_j\psi_k))]+\nonumber\\
&&+\Phi_{xx}[\frac{1}{6}\epsilon_{ijk}(\psi_i\psi_j\psi_k-3\lambda_i\lambda_j\psi_k)\psi_0]
+\nonumber\\
&&+\Phi_{yy}[\frac{1}{6}\epsilon_{ijk}(\lambda_i\lambda_j\lambda_k-3\psi_i\psi_j\lambda_k)\lambda_0]+\nonumber\\
&&-\Phi_{xy}[\frac{1}{6}\epsilon_{ijk}(\psi_i\psi_j\psi_k\lambda_0+\lambda_i\lambda_j\lambda_k\psi_0+3\psi_i\lambda_j\lambda_k
\lambda_0+
3\lambda_i\psi_j\psi_k\psi_0)].\nonumber\\
&&
\eea
The action associated to the lagrangian is invariant under $N=8$ off-shell supersymmetries. The presentation above makes use of the manifest, quaternionic-covariant, $N=4$ decomposition into $(1,4,3)+(1,4,3)$ multiplets.
It allows to present the results in terms of the totally antisymmetric $\epsilon_{ijk}$ tensor. 
By setting $x_1=x$  and $x_2=y$, we can express the lagrangian as
${\cal L} = g_{ij}{\dot x}_i{\dot x}_j +\ldots$, where $g_{ij}(x_1,x_2)$ denotes the target manifold metric.
In our case 
\bea
g_{ij} &=& \delta_{ij}\Phi,
\eea
proving that the target is conformally flat. The constraint (\ref{constr}) implies that the target manifold corresponds to a special geometry of rigid type (see \cite{sp} for details). \par
Some comments are in order. The (\ref{lagr}) lagrangian coincides with the result presented in \cite{bikl} for the $N=8$
$(2,8,6)$ action.
Our analysis is however more complete and convenient for at least three reasons: we have imposed the invariance under $N=5$ generators and proven that the resulting action is automatically invariant under $N=8$
supersymmetry. We investigated the role of the multiplets with different connectivity and proved that the
action obtained from (\ref{lagr}) is invariant under both $N=5A$ and $N=5B$. Finally, our computational scheme allows to present the supersymmetric action directy in terms of the component fields. It is quite a non-trivial and cumbersome task to extract the component fields from the constrained superfields employed in \cite{bikl}.
\section{Conclusions}

This work is the first one in an intended systematic program of constructing and analyzing the properties of off-shell invariant $N$-extended one-dimensional supersymmetric $\sigma$-models.
This project is made possible due to two main reasons. The classification, given in \cite{kt}, of the inequivalent irreducible multiplets of the $N$-extended supersymmetries (the \cite{kt} results also allow to explicitly construct a representative multiplet in each equivalence class) and the computational scheme, here discussed, to produce off-shell invariant actions.\par
We concretely investigated the simplest case (the $(2,8,6)$ multiplets) admitting, for the same field content, inequivalent $N=5$ supersymmetries. The associated $\sigma$-model possesses a two-dimensional target. We proved that requiring an $N=5A$ invariance automatically guarantees an
$N=8$ off-shell invariant action. Similarly, requiring an $N=5B$ invariance automatically guarantees a full $N=8$ invariance. The action, which is therefore invariant under both $N=5$ transformations,
corresponds to a conformally flat $2D$ target with a special metric of rigid type (satisfying the (\ref{constr2}) equation).  \par
There is a common wisdom, originated by scattered results produced by several groups using different methods, which deserves being tested. It states that for $N>4$ linear supersymmetries the target manifold should be conformally flat and the
conformal factor should satisfy a Laplace equation. This is indeed what we obtained here. An explicit proof for all cases is however lacking. This would amount to investigate the remaining $(k,8,8-k)$ multiplets, as well as the length-$4$ multiplets (those admitting fermionic auxiliary fields of mass-dimension $\frac{3}{2}$). The length-$4$ multiplets, classified in \cite{krt} and \cite{kt},
cannot be ``oxidized'' to $N=8$ (they only carry $N=5$, $6$ or $7$ supersymmetries, according to the cases). So far they have never been analyzed in the literature. They can now be investigated with the method here developed.\par
The issue of the $N=5$ invariance implying the $N=8$ invariance is also of utter importance. Our result corroborates (for a totally different and in principle unrelated system) the finding of \cite{bbm} obtained by twisting the $N=2$ Super-Yang-Mills theory. It would be important to verify whether this is a general property, valid for any $(k,8,8-k)$ multiplet. \par
Our computational scheme is not limited to $N=8$. It works in principle for $N>8$. Some special cases are of
particular importance. The dimensional reduction of the $N=4$ Super-Yang-Mills theory to $0+1$ dimensions produces \cite{top,{N9}} an $N=9$ off-shell invariance realized on a $(9,16,7)$ multiplet
which can be decomposed, under the $N=4$ subalgebra, into a single $(3,4,1)$ vector multiplet and three $(2,4,2)$
matter multiplets. In terms of the $N=8$ decomposition we end up with the $(5,8,3)$ and $(4,8,4)$
multiplets linked together by a $9$-th supersymmetry \cite{kt2}.\par
The results on twist suggest \cite{N9} that $6$ generators are sufficient to fully determine the
full $N=9$ invariance. This would amount to use $5$ out of the $8$ generators in the $N=8$ decomposition. The remaining generator to be used would be the $9$-th supersymmetry generator.\par
Our computational scheme can allow to attack this one as well as similarly related problems. It is worth mentioning that in the literature no systematic investigation of off-shell invariant actions for $N>8$ have been carried on.

{}~
\\{}~
{\Large{\bf Appendix}}
~\\
~

The $N=5A,5B, 6A, 6B, 7,8$ extended supersymmetries represented on the multiplets with $(2,8,6)$ field-content can be explicitly given as
\bea
N=5A&:& {\widehat Q}_1,{\widehat Q}_2,{\widehat Q}_3,{\widehat Q}_4,
{\widehat Q}_5\nonumber\\
N=5B&:& {\overline Q}_1,{\overline Q}_2,{\overline Q}_3,{\overline Q}_4,
{\overline Q}_5\nonumber\\
N=6A&:& {\widehat Q}_1,{\widehat Q}_2,{\widehat Q}_3,{\widehat Q}_4,
{\widehat Q}_5, {\widehat Q}_6
\nonumber\\
N=6B&:& {\overline Q}_1,{\overline Q}_2,{\overline Q}_3,{\overline Q}_4,
{\overline Q}_5, {\overline Q}_6\equiv {\widehat Q}_1,{\widehat Q}_2,
{\widehat Q}_3,{\widehat Q}_4,{\widehat Q}_5, {\widehat Q}_7\nonumber\\
N=7&:& {\overline Q}_1,{\overline Q}_2,{\overline Q}_3,{\overline Q}_4,
{\overline Q}_5, {\overline Q}_6,{\overline Q}_7\equiv {\widehat Q}_1,{\widehat Q}_2,
{\widehat Q}_3,{\widehat Q}_4,{\widehat Q}_5, {\widehat Q}_6, {\widehat Q}_7\nonumber\\
N=8&:& {\overline Q}_1,{\overline Q}_2,{\overline Q}_3,{\overline Q}_4,
{\overline Q}_5, {\overline Q}_6,{\overline Q}_7,{\overline Q}_8\equiv {\widehat Q}_1,{\widehat Q}_2,
{\widehat Q}_3,{\widehat Q}_4,{\widehat Q}_5, {\widehat Q}_6,{\widehat Q}_7,{\widehat Q}_8
\eea
where the ${\widehat Q}_i$ and ${\overline Q}_i$ generators are expressed in the tables below.
The $N=5A$ $(2,8,6)_A$ multiplet admits $2_5+2_4+4_3$ connectivity (see \cite{kt}) and $N=4$ decomposition
(see \cite{kt2}) given by $(2,4,2)+(0,4,4)$. Its component fields can be parametrized as $(x,y;\psi_1,\ldots, \psi_8; g_1, \ldots, g_6)$. The $N=5B$ $(2,8,6)_B$ multiplet admits $6_4+4_3$ connectivity (see \cite{kt}) 
and $N=4$ decomposition
given by $(1,4,3)+(1,4,3)$ (see \cite{kt2}). It is convenient to parametrize its component fields as
$(x,y;\psi_0,\psi_1,\psi_2,\psi_3,\lambda_0,\lambda_1,\lambda_2,\lambda_3;g_1,g_2,g_3,f_1,f_2,f_3)$
to make its quaternionic structure clear ($g_i, f_i$ are the auxiliary fields). The $N=6A$ multiplet 
admits
$2_6+6_4$ connectivity, while $N=6B$ admits $4_5+4_4$ connectivity. Note that, according to 
the ``oxidation diagram'' of \cite{kt2}, $N=5B$ can only be lifted to the $B$-type $N=6$ supersymmetry, 
while $N=5A$ can be lifted to both $N=6A$ and $N=6B$. \par
The ${\widehat Q}_i$ and ${\overline Q}_i$ 
generators are explicitly given by
{{ { {{\begin{eqnarray}&\label{N5A8}
\begin{tabular}{|c||c|c|c|c|c|c|c|c|}\hline
 &${\widehat Q_{1}}$ &${\widehat Q_{2}}$&${\widehat Q_{3}}$&${\widehat Q_{4}}$& 
 ${\widehat Q_{5}}$&${\widehat Q_{6}}$&${\widehat Q_{7}}$&${\widehat Q_{8}}$\\ \hline\hline
$x$&$\psi_{1}$&$-\psi_{2}$&$-\psi_{3}$&$-\psi_{4}$&$\psi_5$&$\psi_6$&$\psi_7$&$\psi_8$\\
$y$&$\psi_{2}$&$\psi_{1}$&$\psi_{4}$&$-\psi_{3}$&$\psi_{6}$&$-\psi_{5}$&$\psi_{8}$&
$-\psi_{7}$\\\hline
$\psi_{1}$&$\dot{x}$&$\dot{y}$&$g_{1}$&$g_{2}$&$-g_3$&$-g_4$&$-g_5$ &$-g_6$\\
$\psi_{2}$&$\dot{y}$&$-\dot{x}$&$-g_{2}$&$g_{1}$&$-g_{4}$&$g_3$&$-g_6$&$g_5$\\
$\psi_{3}$&$g_{1}$&$g_{2}$&$-\dot{x}$&$-\dot{y}$&$-g_5$&$g_6$&$g_3$&$-g_4$\\
$\psi_{4}$&$g_{2}$&$-g_1$&$\dot{y}$&$-\dot{x}$&$-g_6$&$-g_5$&$g_4$&$g_3$\\
$\psi_{5}$&$g_{3}$&$-g_{4}$&$-g_{5}$&$-g_6$&$\dot{x}$&$-\dot{y}$&$-g_{1}$&$-g_{2}$\\
$\psi_{6}$&$g_4$&$g_3$&$g_6$&$-g_5$&$\dot{y}$&$\dot{x}$&$-g_{2}$&$g_{1}$\\
$\psi_{7}$&$g_{5}$&$-g_{6}$&$g_3$&$g_4$&$g_{1}$&$g_{2}$&$\dot{x}$&$-\dot{y}$\\
$\psi_{8}$&$g_{6}$&$g_5$&$-g_4$&$g_{3}$&$g_{2}$&$-g_{1}$&$\dot{y}$&$\dot{x}$\\\hline
$g_{1}$&$\dot{\psi}_{3}$&$-\dot{\psi}_{4}$&$\dot{\psi}_{1}$&$\dot{\psi}_{2}$&
$\dot{\psi}_{7}$&$-\dot{\psi}_{8}$&$-\dot{\psi}_{5}$&$\dot{\psi}_{6}$\\
$g_{2}$&$\dot{\psi}_{4}$&$\dot{\psi}_{3}$&$-\dot{\psi}_{2}$&$\dot{\psi}_{1}$&
$\dot{\psi}_{8}$&$\dot{\psi}_{7}$&$-\dot{\psi}_{6}$&$-\dot{\psi}_{5}$\\
$g_{3}$&$\dot{\psi}_{5}$&$\dot{\psi}_{6}$&$\dot{\psi}_{7}$&$\dot{\psi}_{8}$&
$-\dot{\psi_{1}}$&$\dot{\psi}_{2}$&$\dot{\psi}_{3}$&$\dot{\psi}_{4}$\\
$g_{4}$&$\dot{\psi}_{6}$&$-\dot{\psi}_{5}$&$-\dot{\psi}_{8}$&$\dot{\psi}_{7}$&
$-\dot{\psi}_{2}$&$-\dot{\psi}_{1}$&$\dot{\psi}_{4}$&$-\dot{\psi}_{3}$\\
$g_{5}$&$\dot{\psi}_{7}$&$\dot{\psi}_{8}$&$-\dot{\psi}_{5}$&$-\dot{\psi}_{6}$&
$-\dot{\psi}_{3}$&$-\dot{\psi}_{4}$&$-\dot{\psi}_{1}$&$\dot{\psi}_{2}$\\
$g_{6}$&$\dot{\psi}_{8}$&$-\dot{\psi}_{7}$&$\dot{\psi}_{6}$&$-\dot{\psi}_{5}$&
$-\dot{\psi}_{4}$&$\dot{\psi}_{3}$&$-\dot{\psi}_{2}$&$-\dot{\psi}_{1}$\\\hline
\end{tabular}&\nonumber\\
&&\end{eqnarray}}} } }}
and

\begin{eqnarray}&\label{N5B8}
\begin{tabular}{|c||c|c|c|c|c|c|c|c|}\hline
 &${\overline Q_{1}}$ &${\overline Q_{2}}$&${\overline Q_{3}}$&${\overline Q_{4}}$& 
 ${\overline Q_{5}}$&${\overline Q_{6}}$&${\overline Q_{7}}$&${\overline Q_{8}}$\\ 
 \hline\hline

$x$&$-\psi_{1}$&$-\psi_{2}$&$-\psi_{3}$&$\psi_{0}$&$\lambda_{0}$&$\lambda_{1}$&$
\lambda_{2}$&$\lambda_{3}$\\\hline
$\psi_{0}$&$g_{1}$&$g_{2}$&$g_{3}$&$\dot{x}$&$-\dot{y}$&$-f_{1}$&$-f_{2}$ &$-f_{3}$\\
$\psi_{1}$&$-\dot{x}$&$g_{3}$&$-g_{2}$&$g_{1}$&$-f_{1}$&$\dot{y}$&$f_{3}$&$-f_{2}$\\
$\psi_{2}$&$-g_{3}$&$-{\dot x}$&$g_1$&$g_{2}$&$-f_{2}$&$-f_{3}$&$\dot{y}$&$f_{1}$\\
$\psi_{3}$&$g_{2}$&$-g_1$&$-{\dot x}$&$g_{3}$&$-f_{3}$&$f_{2}$&$-f_{1}$&$\dot{y}$\\
\hline
$g_{1}$&$\dot{\psi}_{0}$&$-\dot{\psi}_{3}$&$\dot{\psi}_{2}$&$\dot{\psi}_{1}$&$
\dot{\lambda}_{1}$&$-\dot{\lambda}_{0}$&$-\dot{\lambda}_{3}$&$\dot{\lambda}_{2}$\\
$g_{2}$&$\dot{\psi}_{3}$&$\dot{\psi}_{0}$&$-\dot{\psi}_{1}$&$\dot{\psi}_{2}$&$
\dot{\lambda}_{2}$&$\dot{\lambda}_{3}$&$-\dot{\lambda}_{0}$&$-\dot{\lambda}_{1}$\\
$g_{3}$&$-\dot{\psi}_{2}$&$\dot{\psi}_{1}$&$\dot{\psi}_{0}$&$\dot{\psi}_{3}$&$
\dot{\lambda_{3}}$&$-\dot{\lambda}_{2}$&$\dot{\lambda}_{1}$&$-\dot{\lambda}_{0}$\\
\hline
$y$&$\lambda_{1}$&$\lambda_{2}$&$\lambda_{3}$&$\lambda_{0}$&$-\psi_{0}$&$\psi_{1}$&
$\psi_{2}$&$\psi_{3}$\\\hline
$\lambda_{0}$&$-f_{1}$&$-f_{2}$&$-f_{3}$&$\dot{y}$&$\dot{x}$&$-g_{1}$&$-g_{2}$&$-g_{3}$\\
$\lambda_{1}$&$\dot{y}$&$-f_{3}$&$f_{2}$&$f_{1}$&$g_{1}$&$\dot{x}$&$g_{3}$&$-g_{2}$\\
$\lambda_{2}$&$f_{3}$&$\dot{y}$&$-f_1$&$f_{2}$&$g_{2}$&$-g_{3}$&$\dot{x}$&$g_{1}$\\
$\lambda_{3}$&$-f_{2}$&$f_1$&$\dot{y}$&$f_{3}$&$g_{3}$&$g_{2}$&$-g_{1}$&$\dot{x}$\\
\hline
$f_{1}$&$-\dot{\lambda}_{0}$&$\dot{\lambda}_{3}$&$-\dot{\lambda}_{2}$&
$\dot{\lambda}_{1}$&$-\dot{\psi}_{1}$&$-\dot{\psi}_{0}$&$-\dot{\psi}_{3}$&
$\dot{\psi}_{2}$\\
$f_{2}$&$-\dot{\lambda}_{3}$&$-\dot{\lambda}_{0}$&$\dot{\lambda}_{1}$&
$\dot{\lambda}_{2}$&$-\dot{\psi}_{2}$&$\dot{\psi}_{3}$&$-\dot{\psi}_{0}$&
$-\dot{\psi}_{1}$\\
$f_{3}$&$\dot{\lambda}_{2}$&$-\dot{\lambda}_{1}$&$-\dot{\lambda}_{0}$&
$\dot{\lambda}_{3}$&$-\dot{\psi}_{3}$&$-\dot{\psi}_{2}$&$\dot{\psi}_{1}$&
$-\dot{\psi}_{0}$\\\hline
\end{tabular}&\nonumber\\
&&\end{eqnarray}
{}~
\\{}~
\par {\large{\bf Acknowledgments}}{} ~\\{}~\par
M.G. receives a CLAF grant. M.R. receives a FAPEMIG grant.
The work has been supported by Edital Universal CNPq, Proc. 472903/2008-0.

\end{document}